\documentclass[12pt]{article}
\usepackage{color,amsmath,amssymb,amsthm,bm,epsf,epsfig,enumerate,alltt,latexsym,graphicx}
\usepackage{mathabx,textcomp}
\newcommand{\E}{\ensuremath{\mathrm{E}}}
\newcommand{\var}{\ensuremath{\mathrm{Var}}}
\newcommand{\pr}{\ensuremath{^{\prime}}}
\newcommand{\id}{\ensuremath{\mathrm{I}}}
\renewcommand{\P}{\ensuremath{\mathbf{P}\!}}
\newcommand{\N}{\ensuremath{\mathbf{N}}}
\renewcommand{\sp}{\ensuremath{\mathrm{sp}}}
\newcommand{\tr}{\ensuremath{\mathrm{tr}}}
\newcommand{\Diag}{\ensuremath{\mathrm{Diag}}}
\newcommand{\imp}{\ensuremath{\Longrightarrow}}
\newcommand{\pmi}{\ensuremath{\Longleftarrow}}
\newcommand{\mbk}{\ensuremath{\mathbb{K}}}
\newtheorem{propn}{Proposition}

\begin{document}
\begin{center}
\begin{Large}
{\bfseries Yates's MWSM SS\\ in the General Linear Model}\end{Large}\\
\vspace{.5cm}by Lynn R. LaMotte\footnote{School of Public Health, LSU Health, New Orleans, LA,  {\tt llamot@lsuhsc.edu}}
\end{center}

\begin{abstract}
In 1934, F. Yates described a sum of squares for testing factor main effects in saturated unbalanced models for effects of two factors.    He claimed no particular properties of this sum of squares other than that it provided an ``efficient estimate of the variance from the A means of the sub-class means... .''  Although it became widely regarded as the gold standard in the two-factor model, its fundamental properties and relations to other sums of squares for the same model were not established until decades later. Its method has not been extended to more general settings. This paper shows how Yates's approach can be extended to construct numerator sums of squares for test statistics for linear hypotheses in general linear models.  It is shown that Yates's sum of squares is equivalent to the restricted model - full model difference in error sum of squares, which in turn is shown to be the unique sum of squares that tests exactly the hypothesis in question.
\end{abstract}

\vspace{.5cm}\noindent{\sc Key Words}: ANOVA, Linear Models, Main Effects

\section{Introduction}
Test statistics for fixed effects in linear models are, for the most part, $F$-statistics of the form $F = (SS/\nu)/MSE$, where $MSE$ is mean squared error, $SS=\bm{y}\pr P \bm{y}$ is a quadratic form in the vector $\bm{y}$ of responses, which will be called here the \emph{numerator sum of squares}, $P$ is a symmetric, idempotent matrix, and $\nu = \tr(P)$ is the numerator degrees of freedom.  These terms are defined below, but it will help to preview them here.

In a seminal paper, Yates (1934) described the ``method of weighted squares of means'' (MWSM) to obtain a numerator sum of squares (SS) for testing main effects of  factor A in unbalanced  models for main effects of  factors A and B and their interaction effects.  He reasoned that, if $\bm{U}\sim\N(\mu\bm{1}_p, \sigma^2D)$, with $D=\Diag(1/w_i)$, all $w_i>0$, then, quoting his equation (A),
\begin{eqnarray}
Q&=& (p-1)s^2 = w_1(u_1-\bar{u})^2 + w_2(u_2-\bar{u})^2 + \cdots \nonumber\\
&=& w_1u_1^2 + w_2u_2^2 + \cdots - (w_1+w_2+ \cdots)\bar{u}^2 \nonumber\\
& & \text{ where } \bar{u}=\frac{w_1u_1 + w_2u_2 + \cdots}{w_1+w_2+ \cdots}\label{Yates Q}
\end{eqnarray}
  ``provides an efficient estimate'' of $(p-1)\sigma^2$ from the realized value $\bm{u}=(u_1,\ldots,u_p)\pr$ of $\bm{U}$. In matrix terms, $Q$ can be expressed as
\begin{equation}Q=\bm{u}\pr(D^{-1}-D^{-1}\bm{1}(\bm{1}\pr D^{-1}\bm{1})^{-1}\bm{1}\pr D^{-1})\bm{u},\label{Q form}\end{equation}
where $\bm{u}=(u_1,\ldots, u_p)\pr$ and  $\bm{1}$ or $\bm{1}_p$ denotes a vector of ones. 

The MWSM  SS for A main effects comes from this expression upon substituting the ``marginal means of the subclass means'' for $u_i$, with corresponding substitutions for the diagonal entries of $D$.  

Yates proffered no further rationale.  He did not invoke a general approach or set of criteria.  For that reason, it is not clear how to develop $Q$ from basics, or how (or whether) it is related to alternatively-developed sums of squares,  or how to extend the MWSM to other settings with multiple factors and covariates. 

General methods to compute SSs for factor effects (main effects, interaction effects) were developed after  Yates's (1934) paper, but they did not extend the MWSM directly. Like the MWSM, they were developed to be ``suitable for desk calculation'' and to avoid ``solving different sets of simultaneous equations for each sum of squares.'' (Federer and Zelen, 1966).  One of the methods described by Paik and Federer (1974) was, like the MWSM, based on the vector of sample cell means, and so it might be regarded as an extension of the MWSM to multiple factors.  It provided SSs for non-saturated models (where some effects are excluded) and settings in which some factor-level combinations contain no observations (empty cells). Its scope did not extend to general linear models that might include in addition covariates and factor-by-covariate interaction effects.

The advent and widespread use of statistical computing packages required general-purpose procedures. 
The MWSM SS became a touchstone for attempts at more general approaches to assess factor effects in general linear models. It was not provided directly, though, because it was not clear how it could be extended beyond two-factor analysis of variance settings.  When SAS introduced Type III estimable functions and SSs four decades later, for example, Goodnight (1976) asserted (or observed, as he gave no justification), ``When no missing cells exist in a factorial model, Type III SS will coincide with Yate's [sic] weighted squares of means technique.''  Closing that paper, he alluded to the possibility of other good approaches, reflecting the thinking at that time that different definitions of effects and SSs could be equally satisfactory:
\begin{quotation}
Perhaps (and just perhaps) we may someday be able to agree on the estimable functions we want to use in any given situation.  If this day ever comes, we can then consolidate the different types of estimable functions (and live happily ever after).
\end{quotation}

Some relations have been established. Searle (1971, p. 371) showed that the MWSM SS tests equality of the A marginal means by showing that its  noncentrality parameter is 0 if and only if the marginal means are all equal.  That can be deduced from (\ref{Q form}) upon substituting the population marginal means for $\bm{u}$.
Searle, Speed, and Henderson (1981, Appendix B) related it directly to least squares by showing ``after some tedious algebra'' that it could be derived from the form $(G\pr\hat{\bm{\eta}})\pr[\var(G\pr\hat{\bm{\eta}})/\sigma^2]^-(G\pr\hat{\bm{\eta}})$ to test H$_0: G\pr\bm{\eta}=\bm{0}$, where $\bm{\eta}$ is the vector of cell means, columns of  $G$  comprise a complete set of contrasts for the factor main effects in question, and $\hat{\bm{\eta}}$ is the vector of cell sample means. Searle (1987, p. 90) quoted the MWSM SS directly as shown in Yates (1934) and then justified that the resulting $F$-statistic ``is a test statistic for'' the hypothesis of equal A marginal means because, if the marginal means are equal, then the MWSM SS is distributed as proportional to a central chi-squared random variable. 

In 1934, a very positive feature of the MWSM SS was that it was an explicit formula that humans could manage.  Today, with statistical computing packages, it should be possible to obtain an appropriate numerator SS in any linear model for  hypotheses based on any set of estimable functions of the parameters of the mean vector. In models that involve effects of combinations of levels of multiple factors, it is widely thought that the SAS Type III SS (see SAS Institute 1978) is a correct numerator SS for testable hypotheses about factor effects. However, proofs are hard to find, and it is not always clear what a ``correct'' numerator SS is.  

This topic, whether and how to test for main effects in models that do not exclude interaction effects, continues to generate much discussion.   See Searle (1994), Macnaughton (1998), Hector et al. (2010), Langsrud (2003), and Smith and Cribbie (2014). The books by Hocking (2013) and Khuri (2010) give detailed and comprehensive treatments of the topic.  Still, there is disagreement and some confusion on several points.  Those will not be resolved here.

This paper may be considered to be both an appreciation of Yates's MWSM and a rumination as to why it was never extended to models for means in general.  This is approached by suggesting a method of extension to general linear models that reduces to the MWSM in the two-factor setting for which Yates developed it.  That general method describes the construction of a \emph{numerator} SS that \emph{tests exactly} (both defined below) H$_0: G\pr\bm{\beta}=\bm{0}$ in the model $X\bm{\beta}$ for the mean vector of a response.  Then it is shown that such a SS is unique, and that it is equivalent to the Restricted Model - Full Model (RMFM) difference in error SS.  This leads to the conclusion that, while extension of the MWSM is possible, and if ease of ``desk calculation'' is not a consideration, it would offer no advantage over the RMFM SS.

\fbox{See Appendix \ref{defns} for definitions and notation used here.}

\section{Numerator Sums of Squares for Estimable Functions}

Let $\bm{Y}$ denote an $n$-variate random variable, with realized value $\bm{y}$, that follows the model $\bm{Y}\sim \N(X\bm{\beta}, \sigma^2\id)$.  $X$ is a given $n\times k$ matrix of constants; $\bm{\beta}$ is an unknown $k$-vector of parameters; and $\sigma^2$ is an unknown positive parameter.   That is, $\bm{Y}$ follows a multivariate normal distribution with mean vector $\E(\bm{Y}) =\bm{\mu}=X\bm{\beta}$, for some $\bm{\beta}\in\Re^k$, and variance-covariance matrix $\var(\bm{Y})=\sigma^2\id$. This is often called the Gauss-Markov model.  The model (the set of possible vectors) for the mean vector is $\{\bm{\mu}=X\bm{\beta}: \bm{\beta}\in\Re^k\} = \sp(X)$. This is the {\itshape full model} in the discussion here.

The least-squares estimate of $\bm{\mu}=X\bm{\beta}$ in $\sp(X)$, which minimizes $(\bm{y}-X\bm{b})\pr(\bm{y}-X\bm{b})$, is $\hat{\bm{\mu}} = \hat{\bm{y}} = \P_X\bm{y}$.  A function $\hat{\bm{\beta}}$ of $\bm{y}$  such that $X\hat{\bm{\beta}} = \P_X\bm{y}$ for all $n$-vectors $\bm{y}$ is called \emph{a least-squares solution}.  Residual, or error, SS is $SSE = (\bm{y}-X\hat{\bm{\beta}})\pr(\bm{y}-X\hat{\bm{\beta}}) = \bm{y}\pr(\id-\P_X)\bm{y}$.  If $\bm{\mu}\in\sp(X)$, then mean squared error, $\hat{\sigma}^2 = MSE = SSE/\nu_E$, with degrees of freedom $\nu_E = \tr(\id-\P_X)$, is an unbiased estimator of the population variance $\sigma^2$. 

For a $k\times c$ matrix $G$, the function $G\pr\bm{\beta}$ is said to be {\itshape estimable} iff $\sp(G)\subset\sp(X\pr)$.  See Seely (1977) for a careful treatment of estimability and its relation to testing linear hypotheses of the form H$_0: G\pr\bm{\beta} = \bm{0}$.  (Non-zero right-hand sides entail no essential complications, but we shall restrict attention here to $\bm{0}$ for simplicity.)

The conventional test statistic for a linear hypothesis takes the form of an $F$-statistic, $F = (SS/\nu)/MSE$.    The \emph{numerator SS} is a quadratic form $\bm{y}\pr P \bm{y}$, where $P$ is a symmetric, idempotent matrix such that $\sp(P)\subset\sp(X)$.  It follows that it is distributed as $\sigma^2$ times a chi-squared random variable with $\nu=\tr(P)$ degrees of freedom.  Its noncentrality parameter is $\delta^2_P = \bm{\beta}\pr X\pr PX\bm{\beta}/\sigma^2$.  
The fact that $\sp(P)\subset\sp(X)$ implies that $P\bm{y} = P\hat{\bm{y}}$, and so the numerator SS is a function only of the least-squares estimator of the mean vector.  As a consequence, the numerator SS and MSE are independent.

We shall say that a numerator SS $\bm{y}\pr P\bm{y}$ \emph{tests} $G\pr\bm{\beta}$ iff $\delta_P^2 = 0$ implies that $G\pr\bm{\beta} = \bm{0}$; and that it \emph{tests exactly} $G\pr\bm{\beta}$ iff in addition $G\pr\bm{\beta}=\bm{0}$ implies that $\delta_P^2 = 0$.  
Thus $\bm{y}\pr P\bm{y}$ tests $G\pr\bm{\beta}$ iff $\sp(X\pr P)^\perp \subset \sp(G)^\perp$, and exactly iff $\sp(X\pr P)^\perp = \sp(G)^\perp$.  As used in this sense, ``tests $G\pr\bm{\beta}$'' may be read as short for ``tests H$_0: G\pr\bm{\beta} = \bm{0}$.''  However, it can be taken more broadly to mean that the statistic is responsive to the condition.  That is, the distribution of $\bm{y}\pr P\bm{y}$ is different depending on whether $G\pr\bm{\beta}=\bm{0}$ or not.

Let $N$ denote a matrix such that $\sp(N)=\{\bm{\beta}\in\Re^k: G\pr\bm{\beta}=\bm{0}\}$.  That is, $\sp(N)=\sp(G)^\perp$.  Under the condition that $G\pr\bm{\beta}=\bm{0}$, the restricted model is $\{X\bm{\beta}: \bm{\beta}\in\Re^k \text{ and } G\pr\bm{\beta}=\bm{0}\} = \sp(XN)$.  

Assuming that $G\pr\bm{\beta}$ is estimable, let $H$ denote a matrix with columns in $\sp(X)$ such that $X\pr H=G$.  Then $\sp(H) = \sp(X)\cap\sp(XN)^\perp$ and $\P_H = \P_X-\P_{XN}$.  [Proof: That $N\pr G=N\pr X\pr H = 0$ implies that $\sp(H)\subset\sp(X)\cap\sp(XN)^\perp$.  If $X\bm{b}\in\sp(X)\cap\sp(XN)^\perp$, then $N\pr X\pr X\bm{b}=\bm{0}$, which implies that $X\pr X\bm{b} = G\bm{c} = X\pr H\bm{c}$ for some $\bm{c}$, and hence $X\bm{b}-H\bm{c}$ is in $\sp(X)\cap\sp(X)^\perp = \{\bm{0}\}$, which implies that $X\bm{b}=H\bm{c}\in\sp(H)$.]

Proposition \ref{ns cond} in the appendix establishes that, if $G\pr\bm{\beta}$ is estimable in the model $X\bm{\beta}$, then there is exactly one numerator SS that tests exactly $G\pr\bm{\beta}$.  It is $\bm{y}\pr(\P_X - \P_{XN})\bm{y}$, the difference in SSE between the restricted model and the full model, the RMFM SS for $G\pr\bm{\beta}$.

\section{Constructing SSs as Variance Estimates}
Yates's equation (A), (\ref{Yates Q}) above, illustrates what was then a common approach to construct a numerator SS for estimable functions $G\pr\bm{\beta}$ in ANOVA settings. It was to identify independent linear statistics (like $u_1, \ldots, u_p$) all having the same mean if $G\pr\bm{\beta}=\bm{0}$, and then to define a weighted sum of squared deviations among them that ``provides an efficient estimate ... of the variance of the individual observations,''  (Yates 1934, p. 56) that is, of $\sigma^2$.  That ``[t]his estimate of the variance may be compared with the estimate of variance from the variation within sub-classes by means of the $z$ test'' (ibid.) (the $z$-statistic was the logarithm of the $F$-statistic) required that the two sums of squares be independent, which followed if the $u_i$s were linear functions of $\hat{\bm{y}}$.

That heuristic, and hence Yates's MWSM SS $Q$, is extended here to the general setting with  $\bm{Y} \sim \N(X\bm{\beta}, \sigma^2\id)$ to construct a numerator SS for a set $G\pr\bm{\beta}$ of estimable functions.  It is shown that this SS is the unique SS that tests exactly $G\pr\bm{\beta}$, and hence that $Q = \bm{y}\pr \P_H\bm{y} = \bm{y}\pr(\P_X - \P_{XN})\bm{y}$, which is the RMFM SS. In the next section it is shown how this plays out in the two-factor ANOVA setting to coincide with Yates's MWSM SS.
 
 That $G\pr\bm{\beta}$ is estimable in the model $X\bm{\beta}$ implies that there exists a matrix $H_0$ such that $X\pr H_0 = G$.  Because $\P_X X = X$ and $\P_X$ is symmetric, $X\pr H_0 = X\pr (\P_X H_0)$, and so there exists a matrix $H = \P_X H_0$ with all its columns in $\sp(X)$ (equivalently, $\sp(H)\subset\sp(X)$) such that $X\pr H = G$.  Further, it can be seen that this $H$ is unique. 
 
Let $A$ and $C$ denote matrices such that $A$ has linearly independent columns in $\sp(X)$ and $AC = H$.  This guarantees that $D=A\pr A$ is positive-definite (pd) and hence has an inverse, and that $A\pr\bm{y} = (\P_XA)\pr\bm{y} = A\pr \P_X\bm{y} = A\pr\hat{\bm{y}}$ is a function of the estimated mean vector.  Entries of $\bm{U} = A\pr\bm{Y}$ in this general setting correspond to Yates's marginal means of the subclass means in the two-factor setting he considered.  

Matrices $A$ and $C$ satisfying these conditions exist in any case; there may be multiple choices of $A$ and $C$, but $AC = H$ is unique.  It is clearly possible to choose $A$ such that $D$ is diagonal, or even the identity matrix.  That may be a convenient choice in some situations, but in others it might require unnecessary computations.  The only requirement here is that $D$ be pd.

With $\var(A\pr\bm{Y}) = D\sigma^2$, let $\bm{Z}=D^{-1/2}\bm{U} = D^{-1/2}A\pr\bm{Y}$, so that $\bm{Z}\sim \N(D^{-1/2}A\pr X\bm{\beta}, \sigma^2\id)$.   
Let $M$ be a matrix such that $\sp(M)=\sp(C)^\perp$.  With the $c$ columns of $A$ linearly independent and in $\sp(X)$, it follows that 
\[\sp(A\pr X) = \sp(A\pr) = \Re^c.\]
 Then 
\begin{eqnarray*}\{A\pr X\bm{\beta}: \bm{\beta}\in\Re^k \text{ and } G\pr\bm{\beta}=\bm{0}\} &=& \{A\pr X\bm{\beta}: \bm{\beta}\in\Re^k \text{ and } C\pr A\pr X\bm{\beta}=\bm{0}\}\\
& =& \{\bm{\theta} \in\Re^c: C\pr\bm{\theta}=\bm{0}\}\\
& =& \sp(C)^\perp = \sp(M).
\end{eqnarray*}
If $\bm{\beta}$ is such that $G\pr\bm{\beta}=\bm{0}$ then $\bm{Z} \sim \N(D^{-1/2} M\bm{\gamma}, \sigma^2\id)$ for some $\bm{\gamma}$: this is the restricted model for $\bm{Z}$ under H$_0$.   Thus $MSE$ in this null model for $\bm{Z}$ is an unbiased estimator of $\sigma^2$ when H$_0: G\pr\bm{\beta}=\bm{0}$ is true.  Residual SS in this model is
\begin{eqnarray} SSE_{\bm{z}} &=& \bm{z}\pr (\id-\P_{D^{-1/2}M})\bm{z}\nonumber\\
&=& \bm{u}\pr (D^{-1} - D^{-1}M(M\pr D^{-1}M)^-M\pr D^{-1})\bm{u}.\label{gen Yates}
\end{eqnarray}
This corresponds to (\ref{Q form}) and is equivalent to $Q$ in the setting that Yates (1934) considered, as shown in the next section.

Note further that
\begin{eqnarray*}
\bm{z}\pr(\id-\P_{D^{-1/2}M})\bm{z} &=& \bm{z}\pr \P_{D^{1/2}C}\bm{z}, \text{ by Proposition \ref{prop1}},\\
&=& \bm{y}\pr AC(C\pr D C)^-C\pr A\pr\bm{y} \\
&=& \bm{y}\pr \P_{AC}\bm{y}\\
&=& \bm{y}\pr \P_H \bm{y}, \text{ because } AC=H,\\
&=& \bm{y}\pr (\P_X - \P_{XN})\bm{y}.
\end{eqnarray*}
Therefore $SSE_{\bm{z}} = \bm{y}\pr \P_H\bm{y} = \bm{y}\pr(\P_X-\P_{XN})\bm{y}$, and hence $SSE_{\bm{z}}$ is the RMFM SS for $G\pr\bm{\beta}$.  It is the unique SS that tests exactly H$_0: G\pr\bm{\beta}=\bm{0}$.

The development here shows that the vaguely-defined approach that Yates followed, find a quadratic form in $\hat{\bm{y}}$ that is an unbiased estimator of $\sigma^2$ if H$_0$ is true, can be extended to a general linear model and estimable functions $G\pr\bm{\beta}$.  The SS that it produces is the RMFM SS, which is the unique SS that tests exactly H$_0$.  Further, although different choices of $A$ are possible, all lead to the same SS, and so there is some room to choose an $A$ that is convenient.

In Yates's formulation, this approach led to a closed-form algebraic expression for the SS, which made it practicable for computations at that time. That does not seem to be such an important consideration today.  It is simpler and more direct to go straight to $\bm{y}\pr \P_H \bm{y}$.

\section{SS for A Main Effects in the Two-Factor ANOVA Model}\label{ANOVA model}
In the two-factor ANOVA model, denote levels of factors A and B by $i$ and $j$, respectively, $i=1,\ldots, a$, $j=1,\ldots, b$; denote the number of observations on the response under each factor-level combination (also called a {\itshape cell}) $i,j$ by $n_{ij}$, and assume that all $n_{ij}>0$ (there are no empty cells). Denote the population cell means of the response by $\eta_{ij}$ and the  $ab$-vector of cell means by $\bm{\eta}$. Let $n_{\cdot\cdot} = \sum_{ij}n_{ij}$.  For each observation $s=1,\ldots, n_{\cdot\cdot}$, define  the $s$-th row of the $n_{\cdot\cdot}\times ab$  matrix $\mbk$ to have  1 in the column corresponding to the factor-level combination $i,j$ under which the $s$-th subject was observed, and all other entries 0.  Then there is exactly one 1 in each row, and, in the $i,j$-th column, there are $n_{ij}$ 1s, $i=1,\ldots, a$, $j=1,\ldots, b$.  

Denote the $n_{\cdot\cdot}$-vector of the responses by $\bm{Y}$ and its realized value by $\bm{y}$.
The model for the mean vector $\bm{\mu}=\E(\bm{Y})$ of the response is $\mbk\bm{\eta}$, corresponding to $X\bm{\beta}$ in the general formulation above.  The columns of $\mbk$ are linearly independent, and so all linear functions of $\bm{\eta}$ are estimable.  

Yates (1934) defined A main effects as differences among the A {\itshape population marginal means} $\bar{\eta}_{i\cdot} = (1/b)\sum_j\eta_{ij}$, $i=1,\ldots, a$. The $a$-vector of A marginal means can be expressed as $\bm{\theta}=(1/b)(\id_a\otimes \bm{1}_b)\pr\bm{\eta}$. 
The hypothesis of equal A marginal means is H$_0: S_a\bm{\theta}=\bm{0}$, or, in terms of $\bm{\eta}$, H$_0: [(1/b)(\id_a\otimes \bm{1}_b) S_a]\pr\bm{\eta}=\bm{0}$. This takes the form H$_0: G\pr\bm{\beta}=\bm{0}$ with $\bm{\beta}=\bm{\eta}$ and $G= (1/b)(\id_a\otimes \bm{1}_b) S_a = (1/b)(S_a\otimes \bm{1}_b)$.

Let $D_{ab}=(\mbk\pr \mbk)^{-1} = \Diag(1/n_{ij})$. 
To express the numerator SS in the form (\ref{gen Yates}),  $X=\mbk$ and $G=X\pr AC$ with 
\[A=(1/b)\mbk D_{ab}(\id_a\otimes \bm{1}_b)\] 
and $C=S_a$. 
Then $M=\bm{1}_a$ so that $\sp(M)=\sp(C)^\perp$.
Then  $\bm{\theta}=A\pr X\bm{\beta} = A\pr \mbk\bm{\eta}=(1/b)(\id_a\otimes\bm{1}_b)\pr\bm{\eta}$ is the $a$-vector of A population marginal means $\bar{\eta}_{i\cdot}$; and 
$\hat{\bm{\theta}}=A\pr\bm{y}=(\bar{\bar{y}}_{i\cdot}=(1/b)\sum_j\bar{y}_{ij})$ is the $a$-vector of averages,  over levels of B, of the sample cell means $\bar{y}_{ij}=\sum_{\ell=1}^{n_{ij}}y_{ij\ell}/n_{ij}$ (which are the $ab$ entries in $\hat{\bm{\eta}}= D_{ab}\mbk\pr\bm{y}$).
Let 
\[ D_a = A\pr A = (1/b^2)(\id_a\otimes\bm{1}_b\pr)D_{ab}(\id_a\otimes\bm{1}_b) = (1/b^2)\Diag\left(\sum_j(1/n_{ij})\right).
\]
Diagonal entries of $D_a$ are $1/w_i$ in (\ref{Q form}).  With these specifications, 
(\ref{gen Yates}) 
is identical to (\ref{Q form}), the MWSM numerator SS for A main effects. By the results in the last section, this is in turn equal to the RMFM SS.  

Defining matrices $A$ and $C$ in this way corresponds to Yates's (1934) formulation. This has the consequence that $M=\bm{1}_a$  is a column vector, which avoids matrix operations in (\ref{gen Yates}). Another possible choice is $A=(1/b)\mbk D_{ab}$, so that $\bm{\theta}=\bm{\eta}$ and $C=S_a\otimes \bm{1}_b$.  That would result in $M$ having at least $ab-(a-1)$ columns. It is an alternative, and it would lead to the same SS, but evaluating (\ref{gen Yates}) appears to be more burdensome than with $A$ as defined above.

\begin{appendix}
\section{Notation, Definitions, and Facts}\label{defns}

``If and only if'' is abbreviated iff.  In the notation shown next, assume for each that the matrix dimensions are such that the operations are defined.  Matrix notation is standard for addition, product, and inverse.  Generalized inverse and transpose of a matrix $A$ are denoted $A^-$ and $A\pr$, and $\tr(A)$ denotes the trace of $A$ if $A$ is square. Concatenation of columns of matrices $A$ and $B$ having the same number of rows is denoted $(A,B)$.  The only unstated assumption about matrices that appear here is that they exist, that is, they have at least one row and one column.  

Vectors here are column vectors, or matrices with one column; they will be denoted in boldface, e.g., $\bm{z}$.  For an $n\times c$ matrix $M$, $\sp(M)$ denotes the linear subspace of real $n$-dimensional Euclidean space $\Re^n$ spanned by the columns of $M$: that is, $\sp(M) = \{M\bm{x}: \bm{x}\in\Re^c\}$. It is often denoted $C(M)$ and called the ``column space'' of $M$. 
Orthogonality of vectors $\bm{u}$ and $\bm{v}$ in $\Re^n$ is defined by $\bm{u}\pr\bm{v}=0$.  
 The orthogonal complement of $\sp(M)$, denoted $\sp(M)^\perp$, is the set of all $n$-vectors that are orthogonal to all the vectors in $\sp(M)$.  

$\P_M$ denotes the orthogonal projection matrix onto $\sp(M)$. It is defined by two conditions: for any $n$-vector $\bm{z}$, $\P_M\bm{z}\in\sp(M)$ and $\bm{z}-\P_M\bm{z} \in \sp(M)^\perp$.  For any generalized inverse $(M\pr M)^-$ of $M\pr M$, $M(M\pr M)^-M\pr = \P_M$.  $\P_M$  can also be computed as $BB\pr$, where columns of $B$ comprise an orthonormal basis for $\sp(M)$, which can be had by applying the Gram-Schmidt algorithm to $M$.  See LaMotte (2014). 
The relation between linear subspaces and their orthogonal projection matrices is one-to-one: $\sp(M_1)=\sp(M_2)$ if and only if $\P_{M_1} = \P_{M_2}$. Orthogonal projection matrices are symmetric and idempotent.

The Kronecker product of $A$ and $B$, denoted $A\otimes B$, is the matrix formed by replacing each entry $a_{ij}$ of $A$ by $a_{ij}B$.  It can be shown that $(A\otimes B)(C\otimes D) = (AC)\otimes(BD)$ if the matrix products $AC$ and $BD$ are defined: this fact is used in Section \ref{ANOVA model}.

  For a positive integer $m$, let $\bm{1}_m$ denote an $m$-vector of ones, $U_m = (1/m)\bm{1}_m\bm{1}_m\pr$, and $S_m = \id_m - U_m$.  For an $m$-vector $\bm{z}$, $U_m\bm{z}$ replaces each entry  in $\bm{z}$ by $\bar{z} = (1/m)\sum_i z_i$, and $S_m\bm{z}$ replaces each entry  $z_i$ by $z_i-\bar{z}$.  $S_m$ and $U_m$ are symmetric and idempotent, and $S_mU_m=0$.

  If $P$ is $n\times n$, symmetric, and idempotent, and if the $n$-variate random variable $\bm{Y}$ follows a multivariate normal distribution with mean vector $
E(\bm{Y})=\bm{\mu}$ and variance-covariance matrix $\sigma^2\id$ (signified as $\bm{Y} \sim \N_n(\bm{\mu}, \sigma^2\id_n))$, then $\bm{Y}\pr P\bm{Y}/\sigma^2 \sim \chi_\nu^2(\delta^2)$, with  $\nu=\tr(P)$ degrees of freedom and noncentrality parameter $\delta^2 = \bm{\mu}\pr P\bm{\mu}/\sigma^2$.

\begin{propn}
Let $R$ be an $r\times c$ matrix, $M$ a matrix such that $\sp(M)=\sp(R)^\perp$, $D$ an $r\times r$ symmetric positive-definite (pd) matrix, $D^{1/2}$ a symmetric pd matrix such that $D^{1/2}D^{1/2}=D$, and $D^{-1/2}=(D^{1/2})^{-1}$. Then
\[  \P_{D^{1/2}R} = \id - \P_{D^{-1/2}M}.
\]\label{prop1}
\end{propn}

\noindent Proof is left to the reader.

\begin{propn}
Let $X$ be an $n\times k$ matrix.  Let $G$ be a matrix such that $\sp(G)\subset\sp(X\pr)$, and let $N$ be a matrix such that $\sp(N)=\sp(G)^\perp$. If $P$ is a symmetric idempotent matrix such that $\sp(P)\subset\sp(X)$ , then $\sp(X\pr P)^\perp = \sp(G)^\perp$ iff $P=\P_X - \P_{XN}$.\label{ns cond}
\end{propn}

\vspace{.3cm}{\bfseries Proof.}  \fbox{$\imp$:} That $\sp(X\pr P)=\sp(N)^\perp$ $\imp$ $(XN)\pr P = 0$ $\imp$ $\P_{XN}P = 0$ $\imp$ $(\P_X-\P_{XN})P=\P_XP$, and, since $\sp(P)\subset\sp(X)$, $\P_XP=P$.  Therefore $\sp(P)\subset\sp(\P_X-\P_{XN})$.

That $\bm{z}\in\sp(\P_X-\P_{XN})$ $\imp$ $N\pr X\pr \bm{z} = \bm{0}$ $\imp$ $X\pr\bm{z}\in\sp(N)^\perp = \sp(X\pr P)$ $\imp$ $\exists$ $\bm{u}$ such that $X\pr\bm{z} = X\pr P\bm{u}$. With both $\bm{z}$ and $P\bm{u}$ in $\sp(X)$, this implies that $\bm{z}=P\bm{u} \in\sp(P)$. Therefore $\sp(\P_X-\P_{XN}) \subset\sp(P)$. Therefore $\sp(P) = \sp(\P_X-\P_{XN})$, and, because both $P$ and $\P_X-\P_{XN}$ are orthogonal projection matrices onto the same linear subspace, it follows that $P= \P_X -\P_{XN}$. 

\fbox{$\pmi$:} Suppose  $P=\P_X-\P_{XN}$.  
If $\bm{\beta}\in\sp(G)^\perp$ then $\exists$ $\bm{\gamma}$ such that $\bm{\beta}=N\bm{\gamma}$.  Then $PX\bm{\beta} = (\P_X-\P_{XN})XN\bm{\gamma} = \bm{0}$, which implies that $\sp(G)^\perp \subset\sp(X\pr P)^\perp$.  

If $\bm{\beta}\in\sp(X\pr P)^\perp$, then  $PX\bm{\beta} = \bm{0}$ $\imp$ $(\P_X-\P_{XN})X\bm{\beta} = \bm{0}$ $\imp$ $X\bm{\beta}=\P_{XN}X\bm{\beta} = XN\bm{\gamma}$ for some $\bm{\gamma}$. Because $\sp(G)\subset\sp(X\pr)$, $\exists$ $H$ such that $G=X\pr H$.  Then $G\pr\bm{\beta} = H\pr X\bm{\beta} = H\pr X N\bm{\gamma} = G\pr N\bm{\gamma} = \bm{0}$ because $\sp(N)=\sp(G)^\perp$.  Therefore $\sp(X\pr P)^\perp \subset\sp(G)^\perp$.  Therefore $\sp(G)^\perp = \sp(X\pr P)^\perp$.\hfill$\square$ 

\end{appendix}

\section{Bibliography}
\begin{itemize}\setlength{\topsep}{0cm}\setlength{\labelsep}{0cm}
\setlength{\itemsep}{0cm}\setlength{\parsep}{0cm}
\setlength{\parskip}{0cm}\setlength{\itemindent}{-1cm}
%\item[] Anderson, R. L., and Bancroft, T. A. (1952).  Statistical Theory in Research.  McGraw-Hill Book Company, New York.
\item[] Federer, W. T., Zelen, M. (1966).  Analysis of multifactor classifications with unequal numbers of observations.  Biometrics 22(3): 525-552.
\item[] Goodnight, J. H. (1976). The General Linear Models procedure. Proceedings of the First International SAS User's Group.  SAS Institute Inc., Cary, NC.
\item[] Hector, A., von Felten, S., and Schmid, B. (2010).  Analysis of variance with unbalanced data: an update for ecology \& evolution.  Journal of Animal Ecology 79: 308-316.
%\item[] Herr, D. G. (1986).  On the history of ANOVA in unbalanced, factorial designs: the first 30 years.  The American Statistician 40: 265-270.
\item[] Hocking, R. R. (2013). Methods and Applications of Linear Models, Third Edition.  John Wiley \& Sons, Inc., Hoboken, New Jersey.
\item[] Khuri, A. I. (2010). Linear Model Methodology.  Chapman \& Hall/CRC, Boca Raton, FL.
\item[] LaMotte, L. R. (2014). The Gram-Schmidt construction as a basis for linear models.  The American Statistician 68: 52-55.
\item[] Langsrud, \O. (2003).  ANOVA for unbalanced data: Use Type II instead of Type III sums of squares.  Statistics and Computing 13:163-167.
\item[] Macnaughton, D. B. (1998).  Which sums of squares are best in unbalanced analysis of variance? MatStat Research Consulting Inc.
\item[] Paik, U. B., Federer, W. T. (1974). Analysis of nonorthogonal $n$-way classifications.  Annals of Statistics 2(5): 1000-1021.
%\item[] Pearson, K. (1900). On the criterion that a given set of deviations from the probable in the case of a correlated system of variables is such that it can be reasonably supposed to have arisen from random sampling. Philosophical Magazine, Series 5, 50: 157-172.
\item[] SAS Institute Inc. (1978).  SAS Technical Report R-101, Tests of hypotheses in fixed-effects linear models.  Cary, NC.
\item[] Searle, S. R. (1971).  Linear Models.  John Wiley \& Sons, Inc., New York.
\item[] Searle, S. R. (1987).  Linear Models for Unbalanced Data.  John Wiley \& Sons, Inc., New York.
\item[] Searle, S. R. (1994).  Analysis of variance computing package output for unbalanced data from fised-effects models with nested factors.  The American Statistician 48: 148-153.
\item[] Searle, S. R., Speed, F. M., and Henderson, H. V. (1981). Some computational and model equivalences in analyses of variance of unequal-subclass-numbers data.  The American Statistician 35: 16-33.
\item[] Seely, J. (1977). Estimability and linear hypotheses. The American Statistician 31: 121-123.
\item[] Smith, C. E., and Cribbie, R.  (2014).  Factorial ANOVA with unbalanced data: A fresh look at the types of sums of squares.  Journal of Data Science 12: 385-404.
%\item[] Snedecor, G. W., and Cox, G. M. (1935). Disproportionate subclass numbers in tables of multiple classification.  Research Bulletin No. 180, Agricultural Experiment Station, Iowa State College of Agriculture and Mechanic Arts, Ames, Iowa.
%\item[] Yates, F. (1933). The principles of orthogonality and confounding in replicated experiments.  The Journal of Agricultural Science 23:108-145.
\item[] Yates, F. (1934).  The analysis of multiple classificatioins with unequal numbers in the different classes.  Journal of the American Statistical Association, 29(185): 51-66.
\end{itemize}

\end{document}